\documentclass[aps,prl,twocolumn,groupedaddress,showpacs]{revtex4}

\usepackage{graphicx}

\begin{document}
\title{Topological vortex formation in a Bose-Einstein condensate}

\author{A.E. Leanhardt}
\author{A. G\"orlitz}
\author{A.P. Chikkatur}
\author{D. Kielpinski}
\author{Y. Shin}
\author{D.E. Pritchard}
\author{W. Ketterle}

\homepage[URL: ]{http://cua.mit.edu/ketterle_group/}

\affiliation{Department of Physics, MIT-Harvard Center for
Ultracold Atoms, and Research Laboratory of Electronics,
Massachusetts Institute of Technology, Cambridge, Massachusetts,
02139}

\date{\today}

\begin{abstract}

Vortices were imprinted in a Bose-Einstein condensate using
topological phases.  Sodium condensates held in a Ioffe-Pritchard
magnetic trap were transformed from a non-rotating state to one
with quantized circulation by adiabatically inverting the magnetic
bias field along the trap axis. Using surface wave spectroscopy,
the axial angular momentum per particle of the vortex states was
found to be consistent with $2\hbar$ or $4\hbar$, depending on the
hyperfine state of the condensate.

\end{abstract}


\maketitle

As superfluids, Bose-Einstein condensates support rotational flow
only through quantized vortices.  The atomic velocity field is
proportional to the gradient of the phase associated with the
macroscopic wavefunction.  This phase winds through an integer
multiple of $2\pi$ around a vortex line.  Such a phase winding can
be imprinted onto the condensate wavefunction either dynamically
or topologically.  Dynamically, the phase of the condensate
evolves according to the time integral of its energy, which can be
tailored locally with a spatially varying external potential.
Topologically, the phase of the condensate advances through
adiabatic variations in the parameters of the Hamiltonian
governing the system.  This phase, which is solely a function of
the path traversed by the system in the parameter space of the
Hamiltonian, is known as Berry's phase~\cite{BER84}.

In this Letter, we implement the proposal of
Refs.~\cite{NIM00,INO00,OMN02,MMN02} and demonstrate the use of
topological phases to imprint vortices in a gaseous Bose-Einstein
condensate (BEC).  Previously, vortices have been generated in
two-component condensates using a dynamical phase-imprinting
technique~\cite{MAH99} and in single-component condensates by
rotating the cloud with an anisotropic
potential~\cite{MCW00,ARV01,HCE01,HHH02vortex}, by slicing through
the cloud with a perturbation above the critical velocity of the
condensate~\cite{ARV01,IGR01}, and through the decay of
solitons~\cite{DBS01,AHR01}. In this work, $^{23}$Na condensates
were prepared in either the lower, $|F,m_F\rangle = |1,-1\rangle$,
or upper, $|2,+2\rangle$, hyperfine state and confined in a
Ioffe-Pritchard magnetic trap. Vortices were created by
adiabatically inverting the magnetic bias field along the trap
axis and could be removed by returning the bias field to its
original direction. Using surface wave
spectroscopy~\cite{ZAS98,CMD00,HAC01}, we measured the axial
angular momentum per particle of the $|1,-1\rangle$ and
$|2,+2\rangle$ vortex states to be consistent with $-2m_F\hbar$ as
predicted~\cite{NIM00,INO00,OMN02,MMN02}.

\begin{figure}
\begin{center}
\includegraphics{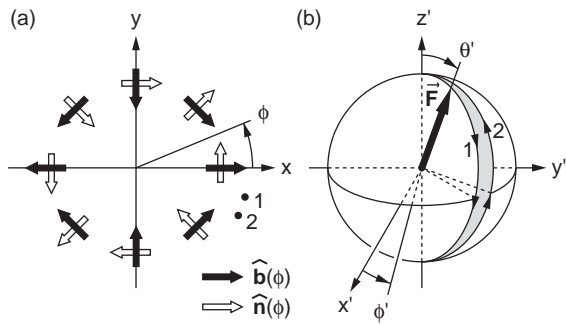}
\caption{Geometry of the rotating magnetic field for the
topological vortex formation.  (a) The unit vectors
$\hat{b}(\phi)$ (solid arrows) point in the direction of the
two-dimensional quadrupole field providing the radial confinement
of a Ioffe-Pritchard magnetic trap. The atomic angular momenta
rotate about the unit vectors $\hat{n}(\phi)$ (open arrows) as the
axial bias field, $B_z$, is ramped from positive to negative
values. (b) For an atom in state $|F,m_F\rangle$, its atomic
angular momentum, $\vec{F}$, traverses a path on a sphere of
radius $|m_F|\hbar$ as it adiabatically follows its local magnetic
field. The primed coordinate system is centered on the atomic
position and has axes parallel to those of the unprimed coordinate
system in (a). For an atomic position described by the azimuthal
angle $\phi$, $\vec{F}$ rotates in a half-plane defined by $\phi'
= -\phi$ for $m_F > 0$ and $\phi' = -\phi + \pi$ for $m_F < 0$ as
$B_z$ is inverted. After inverting $B_z$, the relative topological
phase acquired between atoms located at positions 1 and 2 in (a)
is proportional to the solid angle subtended by the shaded
surface, bounded by the contour marked with arrowheads (see
text).\label{f:berry}}
\end{center}
\end{figure}

A Ioffe-Pritchard magnetic trap consists of an axial bias field
(with curvature) and a two-dimensional quadrupole field in the
orthogonal plane~\cite{GIT62,DEP83,KDS99}:
\begin{equation}
\label{e:noquadratic}
    \vec{B}(x,y,z) = B_z \hat{z} + B' (x \hat{x} - y \hat{y}),
\end{equation}
where $B'$ is the radial magnetic field gradient and quadratic
terms have been neglected.  For a condensate of radial extent $R$,
inverting $B_z$ from $B_z \gg B'R > 0$ to $B_z \ll -B'R < 0$
rotates the atomic angular momentum, $\vec{F}$, by $\pi$ radians.
While all atomic angular momenta rotate through the same angle, a
relative phase is established across the condensate because the
angular momenta rotate about a unit vector $\hat{n}(\phi) = \sin
\phi\ \hat{x} + \cos \phi\ \hat{y}$ that depends on the azimuthal
angle, $\phi$, describing the atomic position
(Fig.~\ref{f:berry}(a)).

As $B_z$ is inverted, $\vec{F}$ adiabatically follows
$\vec{B}(x,y,z)$ and the condensate always remains in the state
$|F,m_F\rangle$ with respect to the local magnetic field. However,
in a basis fixed in the lab frame, the condensate makes the
transition $|F,m_z=+m_F\rangle \rightarrow |F,m_z=-m_F\rangle$,
where $m_F$ and $m_z$ are the projection of $\vec{F}$ along the
local magnetic field direction and z-axis respectively.  Applying
the quantum mechanical rotation operator gives the condensate
wavefunction in the lab frame after inverting $B_z$ as
\begin{eqnarray}
\label{e:rot}
    |\psi\rangle & = & e^{-i \frac{\vec{\mathcal{F}}}{\hbar} \cdot \hat{n}(\phi) \pi} \sqrt{\rho(\vec{r})} |F,m_z=+m_F\rangle,\\
\label{e:vortex}
                 & = & (-1)^{F+m_F} \sqrt{\rho(\vec{r})} e^{-i 2 m_F \phi}|F,m_z=-m_F\rangle,
\end{eqnarray}
where $\vec{\mathcal{F}}$ is the angular momentum operator such
that $\vec{F} = \langle \vec{\mathcal{F}} \rangle$ and
$\rho(\vec{r})$ is the number density of condensed atoms. The
topological phase factor $e^{-i 2 m_F \phi}$ describes a vortex of
winding number $2|m_F|$ with the sense of rotation dependent on
the sign of $m_F$.

This result can be interpreted in terms of Berry's
phase~\cite{OMN02}. Fig.~\ref{f:berry}(b) shows the orientation of
$\vec{F}$ in the lab frame for $m_F > 0$.  Atoms located at
position $k=1,2$ in Fig.~\ref{f:berry}(a) have azimuthal angle
$\phi_k$ and angular momentum $\vec{F}_k$.  As $B_z$ is inverted,
$\vec{F}_k$ traces path $k$ from top to bottom on the sphere in
Fig.~\ref{f:berry}(b).  The topological phase acquired by an atom
in this process is solely a function of the path traversed by its
angular momentum vector. Since this path depends on the azimuthal
angle describing the atomic position, a relative phase is
established between spatially separated atoms.  The condensate
wavefunction after inverting $B_z$ is given by
\begin{equation}
    |\psi\rangle = \sqrt{\rho(\vec{r})} e^{i \gamma(\phi)}
    |F,m_F\rangle,
\end{equation}
where $\gamma(\phi)$ is the topological phase acquired by atoms
with azimuthal angle $\phi$.

For an atom in state $|F,m_F\rangle$, Berry's phase, $\gamma(C)$,
acquired as its angular momentum vector traverses a closed
contour, $C$, on the surface of the sphere in
Fig.~\ref{f:berry}(b) is given by~\cite{BER84}
\begin{equation}
\label{e:berry}
    \gamma(C) = -m_F \Omega(C),
\end{equation}
where $\Omega(C)$ is the solid angle subtended by a surface
bounded by the contour $C$.  The relative phase,
$\gamma(\phi_1)-\gamma(\phi_2)$, is unaltered by closing the
contours traced by each $\vec{F}_k$ along an arbitrary but
identical path.  For clarity, we choose to close each contour
along path 2 itself and hence $\gamma(\phi_1) - \gamma(\phi_2) =
\gamma(C)$, where $C$ is the contour formed by path 1 traversed
from top to bottom and path 2 traversed from bottom to top, as
indicated with arrowheads in Fig.~\ref{f:berry}(b).

A surface bounded by this contour subtends a solid angle
$\Omega(C) = 2(\phi_1 - \phi_2)$, yielding a relative phase
$\gamma(\phi_1) - \gamma(\phi_2) = -2 m_F (\phi_1 - \phi_2)$.
Thus, we make the assignment
\begin{equation}
\label{e:phase}
    \gamma(\phi) = -2 m_F \phi,
\end{equation}
up to an additive term independent of position.  This is the same
phase as in Eq.~\ref{e:vortex} reinterpreted in terms of Berry's
phase.

In this work, Bose-Einstein condensates containing over $10^7$
$^{23}$Na atoms were created in the $|1,-1\rangle$ state in a
magnetic trap, captured in the focus of an optical tweezers laser
beam, and transferred into an auxiliary ``science'' chamber as
described in Ref.~\cite{GCL02}.  While optically confined by the
tweezers, condensates were prepared in the $|2,+2\rangle$ state by
sweeping through the $|1,-1\rangle \leftrightarrow |1,0\rangle
\leftrightarrow |1,+1\rangle$ radio-frequency transition with
100\% efficiency, then sweeping through the $|1,+1\rangle
\leftrightarrow |2,+2\rangle$ microwave transition with 80\%
efficiency~\cite{GGL02}.  In the science chamber, the condensate
was loaded into a microfabricated Ioffe-Pritchard magnetic trap
formed by a Z-shaped wire carrying current $I$ and an external
magnetic bias field, $B_\bot$, as detailed in Ref.~\cite{LCK02}.
Condensates were detected via axial absorption imaging whereby
resonant laser light propagating along the z-axis illuminated the
atoms and was imaged onto a CCD camera~\cite{KDS99}.

Typical wiretrap parameters were $I = 1200$~mA, $B_\bot = 5.4$~G,
and $B_z \approx 1$~G, resulting in a radial magnetic field
gradient of $B' = 120$~G/cm.  For atoms in the $|1,-1\rangle$
state, the axial and radial trap frequencies were $\omega_z = 2
\pi \times 6$~Hz and $\omega_\bot = 2 \pi \times 210$~Hz
respectively. For atoms in the $|2,+2\rangle$ state, the wiretrap
frequencies (for identical magnetic field parameters) were larger
by a factor of $\sqrt{2}$ due to the larger magnetic moment. After
transfer into the wiretrap, condensates in the
$|1,-1\rangle$($|2,+2\rangle$) state contained over $2 \times
10^6$ atoms($1 \times 10^6$ atoms) and had a lifetime in excess of
10~s(3~s) with an applied radio-frequency shield~\cite{KDS99}.
This represents the first magnetic trapping of $^{23}$Na
condensates in the upper hyperfine level, with previous work done
exclusively in optical dipole traps~\cite{GGL02}.

Along the wiretrap axis, the magnetic field is
\begin{equation}
    \vec{B}(x=0,y=0,z) = (B_z + \frac{1}{2}B''z^2) \hat{z},
\end{equation}
where quadratic terms neglected in Eq.~\ref{e:noquadratic} have
been included.  The axial magnetic field curvature, $B''$, which
arises from the geometry of the Z-wire, was held constant
throughout the experiment. By reversing the external axial
magnetic field, we inverted $B_z$.  Changing the sign of $B_z$,
but not $B''$, resulted in a magnetic field saddle point at the
center of the cloud and axial anti-trapping of weak-field seeking
atoms. This limited the condensate lifetime after inverting $B_z$
to $\lesssim 50$~ms.

\begin{figure}
\begin{center}
\includegraphics{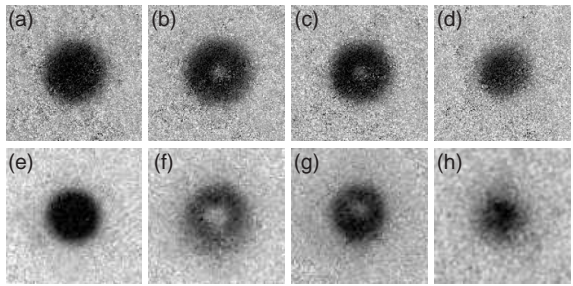}
\caption{Observation of vortices formed by imprinting topological
phases.  Axial absorption images of condensates in the
$|1,-1\rangle$ state after 18~ms of ballistic expansion (a) prior
to inverting $B_z$, after inverting $B_z$ and holding the trapped
condensate for (b) 5~ms and (c) 20~ms, and (d) after inverting
$B_z$ and then returning it to its original direction.  Axial
absorption images of condensates in the $|2,+2\rangle$ state after
7~ms of ballistic expansion (e) prior to inverting $B_z$, after
inverting $B_z$ and holding the trapped condensate for (f) 0~ms
and (g) 5~ms, and (h) after inverting $B_z$ and then returning it
to its original direction. The field-of-view is (a)-(d) 570~$\mu$m
$\times$ 570~$\mu$m and (e)-(h) 285~$\mu$m $\times$ 285~$\mu$m.
\label{f:spin}}
\end{center}
\end{figure}

Vortices created by inverting $B_z$ were identified by
characteristic centrifugal density depletions observed after
ballistic expansion (Fig.~\ref{f:spin}). These vortices could be
removed by returning $B_z$ to its original direction.

For condensates in the $|1,-1\rangle$ state, the best results were
achieved by inverting the axial bias field linearly from $B_z =
860$~mG to $-630$~mG in 11~ms.  For atoms in the $|2,+2\rangle$
state, the optimum ramp time over the same range was 4~ms. The
field inversion process caused an atom loss of $\approx 50$\% due
to non-adiabatic spin-flips as $B_z$ passed through zero, in
reasonable agreement with calculations in
Refs.~\cite{OMN02,MMN02}. The density depletions shown in
Fig.~\ref{f:spin}(b,c,g) were observed after inverting the axial
bias field and holding the trapped condensate for longer than a
radial trap period.  Thus, the atom loss from the center of the
cloud during the field inversion process could not be responsible
for the observed density depletions. Typical
$|1,-1\rangle$($|2,+2\rangle$) condensates after inverting the
axial bias field contained up to $1 \times 10^6$ atoms($0.5 \times
10^6$ atoms) with a Thomas-Fermi radius $R_{TF} = 5.4 \pm 0.2\
\mu$m($R_{TF} = 4.5 \pm 0.2\ \mu$m) in a trap with radial
frequency $\omega_\bot = 2\pi \times 250$~Hz($\omega_\bot = 2\pi
\times 350$~Hz).

The axial angular momentum per particle of the vortex states was
measured using surface wave spectroscopy~\cite{ZAS98,CMD00,HAC01}.
A superposition of counter-rotating ($m_\ell = \pm 2$) quadrupolar
($\ell=2$) surface waves was excited in the condensate by radially
displacing the magnetic trap center for 200~$\mu$s. Here, $\ell$
and $m_\ell$ characterize the angular momentum and its projection
along the z-axis of the quadrupole modes respectively. This
created an elliptical condensate cross-section with time-dependent
eccentricity.  In the absence of a vortex, the $m_\ell = \pm 2$
quadrupole modes are degenerate and the axes of the elliptical
condensate cross-section remain fixed in time. This degeneracy is
lifted by the presence of a vortex, causing the axes to precess in
the direction of the fluid flow. The precession rate,
$\dot{\Theta}$, is given by~\cite{ZAS98,CMD00,HAC01}
\begin{equation}
    \dot{\Theta} = \frac{\langle L_z \rangle}{2 M \langle r_\bot^2 \rangle}\ ,
\end{equation}
where $\langle L_z \rangle$ is the axial angular momentum per
particle characterizing the vortex state, $M$ is the atomic mass,
and $\langle r_\bot^2 \rangle = \langle x^2 + y^2 \rangle$ is the
mean-squared trapped condensate radius with vortices present.

\begin{figure}
\begin{center}
\includegraphics{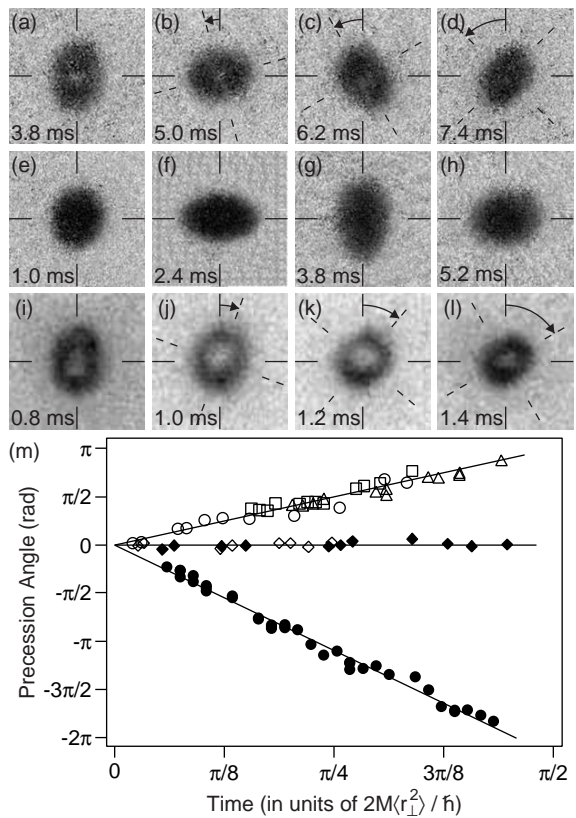}
\caption{Surface wave spectroscopy. Axial absorption images after
18~ms of ballistic expansion of $|1,-1\rangle$ condensates
undergoing a quadrupole oscillation (a)-(d) in the presence of a
vortex and (e)-(h) in the absence of a vortex.  Successive images
were taken during successive half periods of the quadrupole
oscillation such that the short and long axes of the elliptical
cross-section were exchanged. Images (a)-(d) show
counter-clockwise (positive) precession of the quadrupole axes,
while images (e)-(h) show no precession. (i)-(l) Axial absorption
images after 7~ms of ballistic expansion of $|2,+2\rangle$
condensates undergoing a quadrupole oscillation in the presence of
a vortex. The images were taken during a single half period of the
quadrupole oscillation. Images (i)-(l) show clockwise (negative)
precession of the quadrupole axes.  The field-of-view is (a)-(h)
570~$\mu$m $\times$~570 $\mu$m and (i)-(l) 285~$\mu$m $\times$
285~$\mu$m.  (m) Precession angle vs.\ time in the presence of a
vortex for $|1,-1\rangle$ condensates measured after a delay of
0~ms (open circles), 5~ms (open squares), and 20~ms (open
triangles) from the completion of the inversion of the axial bias
field, in the absence of a vortex for $|1,-1\rangle$ (open
diamonds) and $|2,+2\rangle$ (filled diamonds) condensates, and in
the presence of a vortex for $|2,+2\rangle$ condensates measured
immediately upon the completion of the inversion of the axial bias
field (filled circles). \label{f:precession}}
\end{center}
\end{figure}

By measuring the precession rate of the quadrupole axes and the
mean-squared radius of the condensate in the trap, the axial
angular momentum per particle was determined. After exciting the
quadrupolar modes, the condensate evolved in the trap for variable
times in the range $0.2 - 7.4$~ms. The condensate was then
released from the trap and imaged with resonant light after
ballistic expansion as shown in Fig.~\ref{f:precession}(a)-(l).
The resulting images were fit to an elliptical Thomas-Fermi
profile to determine the orientation of the quadrupole axes.  The
orientation angle is plotted as a function of time in
Fig.~\ref{f:precession}(m). To determine the mean-squared trapped
condensate radius, vortices were imprinted in the condensate but
quadrupolar modes were not excited. Images of ballistically
expanded condensates similar to those in
Fig.~\ref{f:spin}(b,c,f,g) were fit to a Thomas-Fermi profile with
circular cross-section. The fitting routine ignored the central
region of the cloud where the density was depleted due to the
vortex core. The mean-squared trapped condensate radius was
derived through the relation
\begin{equation}
    \langle r_\bot^2 \rangle = \frac{2}{7} \frac{R_{\bot}^2}{1+\omega_\bot^2
    \tau^2},
\end{equation}
where $R_{\bot}$ is the Thomas-Fermi radius of the condensate
after ballistically expanding for a time $\tau$ from a trap with
radial frequency $\omega_\bot$.  The factor $1+\omega_\bot^2
\tau^2$ accounts for the change in Thomas-Fermi radius during the
expansion process~\cite{CAD96} and the factor $2/7$ results from
averaging over the inhomogeneous condensate density distribution
assuming no vortices are present.  For low angular momentum vortex
states, the density depletion at the vortex core does not
significantly modify the density distribution of the condensate
and we expect the $2/7$ factor to still be accurate~\cite{DGP99}.

For condensates in the $|1,-1\rangle$ state, the quadrupole
oscillation was excited after a delay of 0, 5, and 20~ms from the
completion of the inversion of the axial bias field. The measured
axial angular momenta per particle were $+1.9(2)(2)\hbar$,
$+2.1(2)(2)\hbar$, and $+1.9(1)(2)\hbar$ respectively, where the
first uncertainty is associated with the linear fit to the
precession angle and the second uncertainty is associated with the
determination of $\langle r_\bot^2 \rangle$. For condensates in
the $|2,+2\rangle$ state, the quadrupole oscillation was excited
immediately upon the completion of the inversion of the axial bias
field.  The measured axial angular momentum per particle was
$-4.4(1)(4)\hbar$.  For both internal states, the measurements are
consistent with the predicted axial angular momentum per particle
of $-2m_F\hbar$~\cite{NIM00,INO00,OMN02,MMN02}.

Multiply charged vortices are unstable against decay into singly
charged vortices~\cite{NOP90}.  From our experiments, we cannot
determine if the condensate contained one multiply charged vortex
or multiple, singly charged vortices.  If multiple vortices were
present, they must be closely spaced since they were not resolved
after ballistic expansion.  Furthermore, if the singly charged
vortices had moved apart considerably, it would have lowered the
extracted value of $\langle L_z \rangle$~\cite{CMD00}, which was
not observed even with delayed probing.

In conclusion, we have used topological phases to imprint vortices
in a Bose-Einstein condensate.  The salient feature of this phase
imprinting technique is the passage of the zero of a magnetic
quadrupole field through the condensate. Confining the condensate
optically would allow for a series of such passages and would lead
to states with very high angular momentum. This technique opens
the potential for studying the stability of multiply charged
vortices and the dynamics of vortex-vortex interactions at short
separations.

We thank J.R.~Abo-Shaeer for a critical reading of the manuscript.
This work was funded by ONR, NSF, ARO, NASA, and the David and
Lucile Packard Foundation. A.E.L.\ acknowledges a graduate
fellowship from NSF.

\end{document}